\documentclass{ws-ijmpa-blank-new}

\usepackage[super,compress]{cite}
\usepackage{graphicx}

\usepackage{amssymb,epsf,latexsym,amssymb,amsmath,mathrsfs}
\usepackage[english]{babel}

\newcommand{\beq}{\begin{equation}}
\newcommand{\eeq}{\end{equation}}
\newcommand{\beqa}{\begin{eqnarray}}
\newcommand{\eeqa}{\end{eqnarray}}
\newcommand{\bsubeqs}{\begin{subequations}}
\newcommand{\esubeqs}{\end{subequations}}

\makeatletter
\@addtoreset{equation}{section}
\renewcommand{\theequation}{\thesection.\arabic{equation}}
\makeatother

\begin{document}
\markboth{F.R. Klinkhamer and T. Mistele}
{Classical stability of higher-derivative $q$-theory ...}

%
\catchline{}{}{}{}{}
%

\title{\vspace*{-20mm}
Classical stability of higher-derivative $q$-theory
in the four-form-field-strength realization}

\author{F.R. Klinkhamer and T. Mistele}

\address{Institute for Theoretical Physics, Karlsruhe Institute of
Technology (KIT),\\ 76128 Karlsruhe, Germany\\
frans.klinkhamer@kit.edu,\;tobias.mistele@student.kit.edu}

\maketitle


\begin{abstract}
We show that higher-derivative $q$-theory in the
four-form-field-strength  realization
does not suffer from the Ostrogradsky instability at the classical level.
\end{abstract}
\vspace*{.0\baselineskip}
{\footnotesize
\vspace*{.25\baselineskip}
\noindent \hspace*{5mm}
\emph{Journal}: \emph{Int. J. Mod. Phys. A} \textbf{32} (2017) 1750090
\vspace*{.25\baselineskip}
\newline
\hspace*{5mm}
\emph{Preprint}:  arXiv:1704.05436  
}
\vspace*{-5mm}\newline
\ccode{PACS Nos.: 95.36.+x, 98.80.Es,45.20.Jj}


\section{Introduction}
\label{sec:introduction}

One particular approach to the long-standing cosmological constant problem \cite{Weinberg1988} is given by 
$q$-theory.\cite{KlinkhamerVolovik2016-static-q,%
KlinkhamerVolovik2016-dynamic-q,KlinkhamerVolovik2016-dyn-cancel-Lambda}
This approach is inspired by condensed matter physics 
and aims to describe the deep quantum vacuum
without knowing the microscopic theory at the high-energy scale.
The approach uses a composite \mbox{(pseudo-)scalar} field $q(x)$
which can be realized in various ways.
In the present paper, we consider the realization
based on a four-form field strength $F$ obtained from
a three-form gauge field $A$~\cite{DuffNieuwenhuizen1980,Aurilia-etal1980}.

A direct kinetic term for the $q$-variable
in the four-form realization has been introduced
in Ref.~\citen{KlinkhamerVolovik2016-q-ball}.
One application of this extended $q$-theory with
an explicit kinetic term for $q$
is the result that perturbations around the constant equilibrium value $q_0$
behave like cold dark matter \cite{KlinkhamerVolovik2016-q-DM,KlinkhamerVolovik2016-more-q-DM}.
The present paper addresses a different, more fundamental issue.
Introducing a kinetic term for $q$ results in a higher-derivative theory
because $q$ itself already contains derivatives of $A$.
Typically, higher-derivative theories are pathological as they suffer from the Ostrogradsky instability \cite{Ostrogradsky1850,Woodard2015}.
The purpose of the present paper is to determine
whether or not the four-form realization of $q$-theory with a direct kinetic term for $q$ is affected by the Ostrogradsky instability.

\section{Theory}
\label{sec:Theory}

In the four-form realization of $q$-theory~\cite{KlinkhamerVolovik2016-static-q,KlinkhamerVolovik2016-dynamic-q}, the pseudoscalar field $q$ is defined by the four-form field strength $F$ associated with the three-form gauge field $A$~\cite{DuffNieuwenhuizen1980,Aurilia-etal1980},
\begin{align}
\label{eq:qfromF}
F_{\alpha \beta \gamma \delta}
\equiv
\nabla_{[\alpha} A_{\beta \gamma \delta]} \,, \quad F_{\alpha \beta \gamma \delta} \equiv q \, \sqrt{-g} \, \epsilon_{\alpha \beta \gamma \delta}\,,
\end{align}
where $ \epsilon_{\alpha \beta \gamma \delta} $ is the Levi-Civita symbol
and $g$ the determinant of the metric $g_{\alpha \beta}$.
We employ the convention $\epsilon_{0123} = -1$,
the metric signature $(- + + + )$, and natural units with $ c = \hbar = 1 $.
For flat spacetime, we use the standard Cartesian coordinates
and the standard Minkowski metric,
\bsubeqs\label{eq:Cartesian-coordinates-Minkowski-metric}
\beqa\label{eq:Cartesian-coordinates}
(x^\alpha)&=&
(x^0,\, x^a)=(x^0,\, x^1,\,  x^2,\, x^3)=(t,\, x,\,  y,\, z)\,,
\\[2mm]
\label{eq:Minkowski-metric}
g_{\alpha \beta}(x) &=&
\eta_{\alpha\beta} \equiv [\text{diag}(-1,1,1,1)]_{\alpha\beta}\,.
\eeqa
\esubeqs
We also define antisymmetrization without normalization factors,
for example,
\mbox{$A_{[\alpha \beta]}$ $\equiv$ $A_{\alpha \beta} - A_{\beta \alpha}$.}
From \eqref{eq:qfromF}, we then obtain $q$ in terms of $A$ and $g$,
\begin{align}
 \label{eq:qfromA}
 q(x) = -\frac{1}{\sqrt{-g(x)}} \, \epsilon^{\alpha \beta \gamma \delta} \partial_\alpha A_{\beta \gamma \delta}(x) \,,  
\end{align}
with the spacetime dependence of the fields shown explicitly.

We consider the action from Ref.~\citen{KlinkhamerVolovik2016-q-ball}, which includes a direct kinetic term for $q$,
\beqa
 \label{eq:action}
  S &=& -\int{d^4x\, \sqrt{-g}\,  \left(\frac{R}{16 \pi G(q)}
  + \frac12 \,C(q)\, \nabla_\beta \, q \, \nabla^\beta q
  + \epsilon(q)\right) }
\nonumber\\[1mm]
&\equiv&
\int dx^0\,  L\left[A,\, \partial_0 A, \partial_0^2 A \right] \,,
\eeqa
where, in the functional of the last expression,
we keep spatial derivatives of $A$ implicit and assume
$A$ without indices to stand
for the collection of all components $A_{\alpha \beta \gamma}$.
We take $G(q)>0$,  $C(q)>0$, and $\epsilon(q)$
in the action  \eqref{eq:action} to be even functions of $q$,
in order to have a manifestly parity-conserving theory.
We further assume $\epsilon(q)$ to be a polynomial in $q^2$,
which is bounded from below ($\epsilon \geq \text{const}$)
and nonconstant ($d\epsilon/d q \ne 0 $).
The gauge fields $A(x)$ are considered to have finite spacetime
support (see Sec.~\ref{subsec:vanishing-A123-gauge} for further discussion).

It is relevant for the analysis below that the action  \eqref{eq:action}
is invariant under gauge transformations of the following form:
\begin{align}
\label{eq:gauge-transformation}
A_{\alpha \beta \gamma}(x) \to   A_{\alpha \beta \gamma}^\prime(x)
= A_{\alpha \beta \gamma}(x)
+\partial_{[\alpha} \lambda_{\beta \gamma]}(x)\,,
\end{align}
for arbitrary (not necessarily infinitesimal)
gauge functions $\lambda_{\beta \gamma}(x)$.

\section{Ostrogradsky instability}
\label{sec:ostrogradsky}

Let us start with a brief review of the Ostrogradsky
instability~\cite{Ostrogradsky1850,Woodard2015}.
To this end, consider a single higher-derivative harmonic oscillator as a model for typical higher-derivative theories,
as discussed in Ref.~\citen{Woodard2015}.
Note that this model suffices to discuss the most important aspects of the Ostrogradsky instability.
In particular, it is not necessary to consider a field-theoretic model, since the results are the same qualitatively.
The reason is that the Ostrogradsky instability is exclusively associated with higher-order time derivatives and not with higher-order spatial 
derivatives \cite{Woodard2015}.

Specifically, take the following Lagrangian~\cite{Woodard2015}
of a higher-derivative harmonic oscillator:
\begin{align}
 \label{eq:lagrangian-ho}
 \overline{L}
 =
 -\frac{\varepsilon}{2}\,
  \frac{m}{\omega^2}\, (\ddot{x})^2
 + \frac{m}{2}\, (\dot{x})^2
 - \frac{m\, \omega^2}{2}\, x^2 \,,
\end{align}
where the overdot stands for the derivative with respect to
time $t$ and where
$ \varepsilon $, $ m $, and $ \omega $ are finite positive parameters.

The equation of motion from
the Lagrangian \eqref{eq:lagrangian-ho}
contains time derivatives of $ x(t) $ up to fourth order,
\beq\label{eq:equation-of-motion-ho}
 \varepsilon\, \frac{m}{\omega^2}\, \ddddot{x}
 + m\,\ddot{x}
 + m\,\omega^2\,x =0\,.
\eeq
Therefore, four initial-data inputs are needed to uniquely
specify a solution.
This implies that there are four canonical variables
which can be chosen as follows:%
\bsubeqs\label{eq:canonical-ho}
\beqa
\overline{Q}_{1} &=& x\,,
\\[2mm]
\overline{P}_{1} &=&
\frac{\partial \overline{L}}{\partial \dot{x} } - \partial_0 \frac{\partial \overline{L}}{\partial \ddot{x}}
= m\,\dot{x} + \varepsilon\, \frac{m}{\omega^2}\, \dddot{x}\,,
\\[2mm]
\overline{Q}_{2} &=& \dot{x} \,,
\\[2mm]
\overline{P}_{2} &=& \frac{\partial \overline{L}}{\partial \ddot{x}} = -\varepsilon\, \frac{m}{\omega^2}\, \ddot{x}\,.
\eeqa
\esubeqs
The canonical Hamiltonian is given by the usual Legendre transformation with respect to these canonical variables,
\beqa
 \label{eq:hamiltonian-ho}
\overline{H}
&=& \overline{P}_{1} \, \dot{\overline{Q}}_{1} + \overline{P}_{2} \, \dot{\overline{Q}}_{2} - \overline{L}
\nonumber\\[1mm]
&=&
 \overline{P}_{1} \, \overline{Q}_{2}
 - \frac{\omega^2}{2\, \varepsilon\, m}\, (\overline{P}_{2})^2
 - \frac{m}{2}\, (\overline{Q}_{2})^2
 + \frac{m\, \omega^2}{2} \, (\overline{Q}_{1})^2 \,.
\eeqa

From expression \eqref{eq:hamiltonian-ho},
it is clear why the higher-derivative harmonic oscillator is unstable.
The Hamiltonian $\overline{H}$ is, namely, linear in the canonical momentum $ \overline{P}_{1} $, which
typically allows for runaway solutions as soon as we add interaction terms to $ \overline{L}$ (for example, a term $-\lambda\,x^4$).
Both the 
$ \overline{P}_{1} \overline{Q}_{2} $ term and the rest of the Hamiltonian $\overline{H}$ can then grow arbitrarily large, while $\overline{H}$ stays constant.
This is how the Ostrogradsky instability reveals itself
at the classical level.

In order to formulate
a quantized theory of the higher-derivative harmonic oscillator, note that the general solution of the classical equation of motion
can be written as
\bsubeqs\label{eq:xho}
\beqa
 x(t)&=&
 \alpha_{+}\, e^{-i k_{+} t}
 +  \alpha_{-}\, e^{i k_{-} t}
 + \alpha_{+}^*\, e^{i k_{+} t}
 + \alpha_{-}^*\, e^{-i k_{-} t}\,,
\\[2mm]
 k_{\pm} &\equiv& \omega \; \sqrt{ \frac{1 \mp \sqrt{1-4\,\varepsilon}}{2\,\varepsilon}}  \,,
\eeqa
\esubeqs
where $ \alpha_{\pm} $ are arbitrary complex numbers.
The Hamiltonian $\overline{H}$ reads~\cite{Woodard2015}
\begin{align}
 \label{eq:Hho}
 \overline{H} = 2\,m\,\sqrt{1-4\,\varepsilon}\,
 \Big( k_{+}^2\, |\alpha_{+}|^2 - k_{-}^2\, |\alpha_{-}|^2 \Big) \,.
\end{align}
These last results
show that quantization can proceed in the usual way by introducing creation and annihilation operators.
There are then two degrees of freedom with opposite energies
as can be seen from the expression \eqref{eq:Hho}.
(A noncanonical quantization scheme~\cite{Woodard2015}
is not considered here, as it leads to problems with unitarity.)
Now suppose that we add interactions to $ \overline{L} $.
Then, positive-energy and negative-energy degrees of freedom will inevitably interact with each other, as $ x $ carries both of them.
Therefore, the vacuum will decay into pairs of
positive-energy and negative-energy degrees of freedom.
This is a manifestation of the Ostrogradsky instability at the quantum level.

\section{Classical stability of higher-derivative $q$-theory}
\label{sec:classical-stability}

As mentioned in the previous section, the Ostrogradsky formalism is sensitive to higher-order time derivatives
but not to higher-order spatial derivatives.
The action \eqref{eq:action}, with a kinetic term
proportional to $ \nabla_\beta \, q \, \nabla^\beta q $,
contains both higher-order time derivatives
and higher-order spatial derivatives.
However, the special form of $q$, namely $ q \propto \epsilon^{\alpha \beta \gamma \delta} \partial_\alpha A_{\beta \gamma \delta} $, implies that the only time derivative in $q$ is the one of the gauge-field
component $ A_{123} $  
(the other gauge-field components $A_{213}$, $A_{231}$, etc.
 are the same as $A_{123}$ up to a factor $\pm 1$).
Therefore, the higher-order time derivatives in \eqref{eq:action} are associated with $ A_{123} $ only.
In short, we have
\begin{align}
\label{eq:lagrangian}
L\left[A,\, \partial_0 A,\, \partial_0^2 A \right]
=
L\left[A,\, \partial_0 A,\, \partial_0^2 A_{123} \right] \,,
\end{align}
where, again, spatial derivatives of $A$ are kept implicit and
$A$ without indices stands
for the collection of all components $A_{\alpha \beta \gamma}$.
We can now proceed in two ways.

\subsection{Gauge choice with nonvanishing $A_{123}$}
\label{subsec:nonvanishing-A123-gauge}

As we are dealing with a gauge theory, we have to fix a gauge before we can
start calculating physical quantities.
To this end, we choose a gauge in which the component $ A_{123} $ does not vanish.
Specifically, our gauge choice is
\beq\label{eq:nonvanishing-A123-gauge}
A_{0 \beta \gamma} = 0\,,
\eeq
leaving $ A_{123} $ as the only nonvanishing component of $A$.
This gauge can be obtained by a gauge transformation
\eqref{eq:gauge-transformation} with the following parameters:
\begin{align}
 \label{eq:gauge-parameters-for-nonvanishing-A123-gauge}
 \lambda_{\beta \gamma}(t, x, y, z) =
 -\frac{1}{2!} \int_{-\infty}^t d\tilde{t} \, A_{0 \beta \gamma}(\tilde{t}, x, y, z) \,,
\end{align}
where the integral runs over a finite interval for gauge fields
$A$ with finite temporal support.
For simplicity, we first neglect gravity, $G(q)=0$.
With the gauge choice \eqref{eq:nonvanishing-A123-gauge}
and flat Minkowski spacetime, expression \eqref{eq:qfromA} gives
\beq\label{eq:q-for-nonvanishing-A123-gauge}
 q = -3! \; \partial_0 A_{123} \,.
\eeq

The first step is now to identify the canonical variables and
to calculate the Hamiltonian in terms of these canonical variables,
with all explicit time derivatives eliminated~\cite{Woodard2015}.
There are two canonical coordinates $ Q_{i} $ with associated momenta $ P_{i} $, for $i= 1, 2$, since the Lagrangian \eqref{eq:lagrangian}
depends on the first and second time derivative of $ A_{123} $.
Introducing factors of $ -3! $ for convenience, we find
\bsubeqs\label{eq:canonical}
\beqa
\label{eq:canonical-Q1}
Q_{1} &=& -3! \, A_{123} \,,
\\[2mm]
\label{eq:canonical-P1}
P_{1} &=& \frac{\partial L}{\partial (-3! \, \partial_0 A_{123} )} - \partial_0 \frac{\partial L}{\partial (-3! \, \partial_0^2 A_{123} )} = \frac{\delta S}{\delta q} \equiv - \mu \,,
\\[2mm]
\label{eq:canonical-Q2}
Q_{2} &=& -3! \, \partial_0 A_{123} = q \,,
\\[2mm]
\label{eq:canonical-P2}
P_{2} &=& \frac{\partial L}{\partial (-3! \, \partial_0^2 A_{123} )} = C(q) \, \partial_0 q \,,
\eeqa
\esubeqs
where $\delta S/\delta q$ in \eqref{eq:canonical-P1} follows from
\eqref{eq:action} with vanishing Ricci curvature scalar, $R=0$.
The quantity $\mu$ defined by \eqref{eq:canonical-P1}
turns out to be a constant, as will be explained below.
The canonical Hamiltonian now reads 
\bsubeqs\label{eq:H}
\beqa
\label{eq:H-QiPi-general}
H &=& \int d^{3}x \;
\Big( P_{1} \, \partial_0 Q_{1}  + P_{2} \, \partial_0 Q_{2} \Big) - L\left[Q_{2},\, P_{2} \right]
\\[1mm]
\label{eq:H-QiPi-specific}
&=& \int d^{3}x \, \left( P_{1}\,Q_{2} + \frac12\, \frac1{C(q)}\, (P_{2})^2 + \frac12\, C(q)\, (\partial_{a} Q_{2})^2 + \epsilon(Q_{2}) \right)
\\
\label{eq:H-QiPi-specific-q}
&=& \int d^{3}x \, \left( \epsilon(q) - \mu \, q + \frac12\, C(q)\, (\partial_0 q)^2 + \frac12\, C(q)\, (\partial_{a} q)^2 \right) \,.
\eeqa
\esubeqs
The Hamiltonian $H$ is conserved and coincides with the energy derived from the gravitational energy-momentum tensor.

We see that the Hamiltonian $H$ from \eqref{eq:H-QiPi-specific}
is linear in the canonical momentum $ P_{1} $, just as
the Hamiltonian $\overline{H}$ of
the higher-derivative harmonic oscillator discussed in
Sec.~\ref{sec:ostrogradsky}. For the present case, however,
the result \eqref{eq:canonical-P1}
and
the $A$ field equations~\cite{KlinkhamerVolovik2016-q-ball}
imply that $ P_{1} $ is constant,
\begin{align}
\label{eq:P1-constant}
\partial_\beta\, P_{1} = \partial_\beta\, \frac{\delta S}{\delta q} = 0 \,.
\end{align}
Therefore, no runaway solutions are possible. More precisely,
the conservation of $ H $ and the result \eqref{eq:P1-constant}
make that $ Q_2 $ and $ P_2 $ do not grow arbitrarily large in time.
In order to see this explicitly, note that the Hamiltonian
\eqref{eq:H-QiPi-specific} is bounded from below for a
fixed constant value of $P_{1}$.
In particular, the terms $P_{1}\,Q_{2}$ and $\epsilon(Q_{2})$
from the integrand of \eqref{eq:H-QiPi-specific}
can be combined into an effective potential
$\epsilon_{\text{eff},P_{1}}(Q_{2})
\equiv \epsilon(Q_{2}) + P_{1} Q_{2}$,
which is a polynomial in $Q_{2}$ and bounded from below.
Recall that, by assumption,
$\epsilon(Q_{2}) $ is a nonconstant polynomial in $(Q_{2})^2$,
which is bounded from below.
Hence, if $Q_{2}$ or $P_{2}$ were to grow arbitrarily large,
this would contradict the conservation of $H$.
In contrast to $Q_{2}$ and $P_{2}$,
the canonical coordinate $ Q_{1} $ is allowed to grow
arbitrarily, as it does not appear in $ H $.
However, no physical quantity will directly depend on $ Q_1 $,
since $ Q_1 \propto A_{123} $ is gauge-noninvariant.
Consequently, the result \eqref{eq:P1-constant}implies
that the linear appearance of $ P_1 $ in $ H $ does not lead to a
dynamical instability.
From \eqref{eq:P1-constant} also follows that the quantity $\mu$
defined by \eqref{eq:canonical-P1} is spacetime independent.

Next, consider the case with standard gravity,  $G(q)=G_N\ne 0$.
Similar arguments as the ones given above show that the Hamiltonian is linear in a single canonical momentum.
It can be shown that this canonical momentum is proportional to
$(1/\sqrt{-g})\, \delta S/\delta q$.
Again, this is exactly what is required to be constant by the equations of motion for $A$, now with gravity present \cite{KlinkhamerVolovik2016-q-ball}.
Therefore, also for the case with gravity, the term of the Hamiltonian with
the linearly appearing canonical momentum can be absorbed into a
well-behaved effective potential, implying that the Ostrogradsky instability is absent.

The absence of the Ostrogradsky instability
for the extended $q$-theory \eqref{eq:action} in
the gauge \eqref{eq:nonvanishing-A123-gauge}
can be illustrated by considering a modified version of the higher-derivative harmonic oscillator discussed in Sec.~\ref{sec:ostrogradsky}.
Consider the modified Lagrangian $ \overline{L}_\text{mod} $ which is obtained from $ \overline{L} $, as given in \eqref{eq:lagrangian-ho},
by omitting the term without time derivatives,
\begin{align}
 \label{eq:lagrangian-ho2}
 \overline{L}_\text{mod} =
 -\frac{\varepsilon}{2}\,\frac{m}{\omega^2}\, (\ddot{x})^2
 + \frac{m}{2}\, (\dot{x})^2 \,.
\end{align}
This modification is motivated by the fact that $ A_{123} $ never appears without a time derivative in the action \eqref{eq:action}, the reason
being gauge invariance. Hence,
we have, for the gauge \eqref{eq:nonvanishing-A123-gauge}, the following
arguments of the Lagrangian:
\begin{align}
\label{eq:lagrangian2}
L\left[A,\, \partial_0 A,\, \partial_0^2 A \right]\,
\Big|^{(\text{gauge}\,A_{0 \beta \gamma} = 0)}
=
L\left[\partial_0 A_{123},\, \partial_0^2 A_{123} \right] \,.
\end{align}

Switching from $ \overline{L} $ to $ \overline{L}_\text{mod} $ has no effect on the canonical variables, since the canonical variables are completely determined by the terms with time derivatives.
As a result, the canonical Hamiltonian $\overline{H}_\text{mod}$
is still linear in the canonical momentum $ \overline{P}_{1} $.
However, the equation of motion derived from $ \overline{L}_\text{mod} $
[given by \eqref{eq:equation-of-motion-ho} without the $m \omega^2 x$ term]
now implies that $ \overline{P}_{1} $ is constant, $ \partial_0 \, \overline{P}_{1} = 0 \, $.
It follows that $\overline{Q}_2$ and $\overline{P}_2$ cannot
grow arbitrarily large in time, since $\overline{H}_\text{mod}$
[given by \eqref{eq:hamiltonian-ho} without the
$ m\, \omega^2\, ( \overline{Q}_{1} )^2 / 2 $ term] is conserved
and, for a fixed constant value of $\overline{P}_{1}$,
bounded from above.
The canonical coordinate  $\overline{Q}_1$ can,
in principle,  grow arbitrarily large, as
$\overline{Q}_1$ without derivatives does not appear in
$\overline{H}_\text{mod}$.
The  Lagrangian $\overline{L}_\text{mod} $
possesses indeed a shift symmetry
[$x(t) \to x'(t) = x(t) + b $ for an arbitrary constant $ b $],
which prevents $ \overline{Q}_1 = x $ from appearing in any physical
quantity. Therefore, the same argument as given above shows that
the modified higher-derivative harmonic oscillator
with Lagrangian \eqref{eq:lagrangian-ho2}
is not affected by the Ostrogradsky instability, as long as
the shift symmetry is imposed.

\subsection{Gauge choice with vanishing $A_{123}$}
\label{subsec:vanishing-A123-gauge}

In order to see what happens if the component $ A_{123} $ associated with the higher-order time derivatives is gauged away, we will now consider an alternative gauge choice. Specifically, take the following
gauge:
\beq\label{eq:vanishing-A123-gauge}
A_{1 \beta \gamma} = 0 \,,
\eeq
which, for flat spacetime, leads to
\beq\label{eq:q-in-vanishing-A123-gauge}
 q = 3! \, \partial_{1} A_{023}
 \,.
\eeq
This gauge can be obtained by a gauge transformation
\eqref{eq:gauge-transformation} with
\begin{align}
 \label{eq:gauge2}
 \lambda_{\beta \gamma}(t, x, y, z) =
 -\frac{1}{2!} \int_{-\infty}^x d\tilde{x}
 \, A_{1 \beta \gamma}(t, \tilde{x}, y, z) \,,
\end{align}
where the integral runs over a finite interval for gauge fields
$A$ with finite spatial support
(a physical context is provided by the $q$-ball solution,
which has $q\approx q_0\ne 0$ in the interior region
and $q=0$ in the exterior region corresponding to absolutely
empty space, as discussed in Ref.~\citen{KlinkhamerVolovik2016-q-ball}).
The possibility to gauge away all higher-order time derivatives
of the Lagrangian
suggests that the theory does not suffer from the Ostrogradsky instability.
We will now explicitly show that this is indeed the case.

As no higher-order time derivatives are associated with $ A_{023} $, there is only one canonical coordinate $ \widetilde{Q} $
with associated momentum $ \widetilde{P} $.
Introducing a factor of $ 3! $ for convenience, we find
\bsubeqs \label{eq:canonical2}
\beqa
 \widetilde{Q} &=& 3! \, A_{023} \,,
 \\[2mm]
 \widetilde{P}  &=& \frac{\partial L}{\partial (3! \, \partial_0 A_{023})} = \partial_{1} \left[\, C(q) \, \nabla^0 q \right] \,.
\eeqa
\esubeqs
The Lagrangian $ L $ depends at least quadratically on both $ \widetilde{Q} $ and $ \widetilde{P} $.
Therefore, the canonical Hamiltonian $ \widetilde{H} $, obtained by a Legendre transformation with respect to $\widetilde{Q}$ and $\widetilde{P}$, will not be linear in any canonical variable.
At this point, we might be tempted to conclude that the theory is not affected by the Ostrogradsky instability, but there is a subtlety.

For flat Minkowski spacetime, we write $ \widetilde{H} $
in terms of $q$ and perform an integration by parts,
\beqa \label{eq:H2}
 \widetilde{H} &=&
 \int d^{3}x \, \widetilde{P} \, \partial_0 \widetilde{Q} - L\left[\widetilde{Q},\, \widetilde{P} \right]
\nonumber\\[1mm]
 &=& \int d^{3}x \, \left( \epsilon(q)
 + \frac12\, C(q)\, (\partial_0 q)^2
 + \frac12\, C(q)\, (\partial_{a} q)^2 \right) \,.
\eeqa
In comparison with the Hamiltonian $H$ from \eqref{eq:H-QiPi-specific-q},
the term $-\mu q$ of the integrand is missing.
For typical nonequilibrium solutions of $ q $-theory, this $- \mu q $ term
generates a nonconstant contribution to $ H $ implying that $ \widetilde{H} $ is not conserved \cite{KlinkhamerVolovik2016-dynamic-q, KlinkhamerVolovik2016-q-DM}.
Here, $ \mu $ is defined as in Sec. \ref{subsec:nonvanishing-A123-gauge},
namely $ \mu \equiv - \delta S / \delta q $,
so that $ \mu $ is constant by the $ A $ field  equations.
In principle, the instability could hide in this nonconservation
of $ \widetilde{H} $.
Furthermore, it is rather unusual that the Hamiltonian depends on the gauge choice.
In electrodynamics, for example, the Hamiltonians derived
in different gauges differ merely
by a total derivative (cf. Sec.~3.5.3 of Ref.~\citen{Maggiore2005}).

Regarding the difference between $H$ and $ \widetilde{H} $, we note that
the $-\mu q$ term is indeed a total derivative in the gauge
with $ A_{1 \beta \gamma} = 0 $.
The reason is that $ \mu $ is constant and $ q \propto \partial_{1} A_{023} \propto \partial_{1} \widetilde{Q} $.
However, as already mentioned in the previous paragraph, typical solutions of $q$-theory are such that the spatial integral over $ q \propto \partial_{1} A_{023} $ does not vanish.
Consequently, $ H $ and $ \widetilde{H} $ do not usually coincide in $q$-theory.
In particular, this implies that $ \widetilde{H} $ differs from the energy
derived from the gravitational energy-momentum tensor, while it does generate the correct time evolution for $ \widetilde{Q} $ and $ \widetilde{P} $.
The discrepancy between $ \widetilde{H} $ and the conserved energy is not
really problematic, as the defining property of the canonical
Hamiltonian is that it generates the time evolution of the canonical variables and not that it is conserved.
It does, however, prevent us from deciding whether or not the theory is
unstable.

We can now solve this problem of the nonconservation of
$ \widetilde{H} $ by simply adding the $ - \mu q $ term,
\beq\label{eq:H2-conserv}
\widetilde{H}_\text{conserved} = \widetilde{H}
- \mu \int d^{3}x \; q \,,
\eeq
with the constant $\mu$ as discussed a few lines below \eqref{eq:H2}.
This addition is allowed since the $-\mu q$ term is a total derivative
according to \eqref{eq:q-in-vanishing-A123-gauge}
and total derivatives do not affect the time evolution of the canonical variables.
In this way, we arrive at a Hamiltonian
which is conserved in flat Minkowski spacetime
and which contains a term linear in the canonical variable
$\widetilde{Q}$, namely the $-\mu q$ term.
But the linear term $-\mu q$ does not lead to an instability, if we
consider this linear term together with the potential term
$\epsilon(q)$ and recall the assumptions on $\epsilon(q)$
as stated in Sec. \ref{sec:Theory}.
(In a general spacetime, $-\mu q$ must be replaced
by $- \sqrt{-g} \, \mu  q $.
Here, $ \mu $ is given
by $-(1/\sqrt{-g})\, \delta S / \delta q $,  
as this combination is constant by the $ A $ field equations
in a general spacetime.
Adding the spatial integral of $ - \sqrt{-g} \, \mu q $
to $ \widetilde{H} $ still amounts to adding the integral of
a total derivative to $ \widetilde{H} $, as $ q $ is proportional to
$ 1/\sqrt{-g} $,
and the same conclusion holds for general spacetimes
as for flat spacetime.)
Hence, also in the gauge \eqref{eq:vanishing-A123-gauge}, we conclude that the Ostrogradsky instability is absent.

\section{Discussion}
\label{sec:discussion}

In the present paper,  
we have shown that the four-form realization of $q$-theory with a kinetic term for $q$ is free from the Ostrogradsky instability.
We have derived this result in two different gauges.
The fact that the ``chemical potential'' $ \mu $
(using the terminology from Ref.~\citen{KlinkhamerVolovik2016-static-q})
is constant has been crucial for both derivations.
With the gauge $ A_{0 \beta \gamma} = 0 $,
a constant $ \mu $ implies that the canonical momentum
appearing linearly in the Hamiltonian is constant.
With  the gauge $ A_{1 \beta \gamma} = 0 $,
the required $-\mu q$ term for a conserved Hamiltonian
is a total derivative only if $ \mu $ is constant.

Note that the treatment up till now has been completely classical.
At the quantum level, the Ostrogradsky instability typically leads to an
additional propagating degree of freedom which carries negative energy,
as discussed in Sec. \ref{sec:ostrogradsky}.
Deciding whether or not a quantized higher-derivative $q$-theory exhibits
the corresponding vacuum-instability problem
requires further study, because we do not yet have a quantized $q$-theory at our disposal.
(It is also possible that the quantized theory requires
a detailed knowledge of all microscopic degrees of freedom.)

Still, we expect the quantized theory corresponding to the action \eqref{eq:action} not to have negative-energy propagating degrees of freedom. The reason is the following.
Consider flat Minkowski spacetime with
the constant equilibrium solution $ q_0 $ for the $q$-field
and the corresponding fixed value $ \mu_0 $ for $ \mu $
(see below for the precise relation).
A linear perturbation $ \phi(x) $ of this constant solution $q_0$ is
defined by
\beq\label{eq:phi}
q(x) \equiv q_0 + \phi(x)/\sqrt{C(q_0)} \,.
\eeq
This scalar field $ \phi(x) $ then satisfies the following Klein--Gordon equation:
\bsubeqs\label{eq:pert}
\beqa
\label{eq:pert-KG-eq}
 \Box \, \phi - m^2(\mu_0)\, \phi  &=& 0 \,,
\\[2mm]
\label{eq:pert-m2}
m^2(\mu_0) &\equiv&
\frac{1}{C(q_0)}\, \frac{d^2 \epsilon}{dq^2}\,\Bigg|_{q=q_0} \,,
\\[2mm]
\label{eq:pert-mu0}
\mu_0 &\equiv&
\frac{d\epsilon}{d q}\,\Bigg|_{q=q_0}\,,
\eeqa
\esubeqs
and the corresponding conserved Hamiltonian can be written as
\begin{align}
 \label{eq:Hpert}
 H = \int d^{3}x \left(
 \frac12\,  (\partial_0 \phi)^2
 + \frac12\,  (\partial_{a} \phi)^2
 + \frac12\, m^2(\mu_0) \, \phi^2
 + \text{const}\right) \,.
\end{align}
Both the linearized equation of motion \eqref{eq:pert-KG-eq}
and the corresponding Hamiltonian \eqref{eq:Hpert}
have precisely the same formal structure as in the case of a fundamental scalar field \cite{KlinkhamerVolovik2016-q-ball,KlinkhamerVolovik2016-q-DM}.

The last observation
suggests that quantizing the four-form realization of
\mbox{$q$-theory} with a kinetic term for $q$ leads to one propagating degree of freedom only, in contrast to what we expect from a typical higher-derivative theory.
If this is indeed correct, and if $ \mu $ remains constant in the quantized theory, we arrive at the following scenario.
Theories suffering from the Ostrogradsky instability typically contain two propagating degrees of freedom with opposite energies, so that
the vacuum can decay into pairs of
positive-energy and negative-energy degrees of freedom.
In our case, however, there is only one propagating degree of freedom.
The additional degree of freedom represented by $ \mu $ is nonpropagating and does not lead to a dynamical instability.

\vspace*{-2mm}
\section*{Acknowledgments}
\noindent
It is a pleasure to thank G.E. Volovik for comments on the manuscript.



\end{document}